\documentclass[aps]{revtex4}
\usepackage{graphicx}

\begin{document}

\title{First-principles Calculation of Superconductivity in Hole-doped LiBC: $T_c=65$ K}

\author{J. K. Dewhurst$^1$, S. Sharma$^1$, C. Ambrosch-Draxl$^1$, B. Johansson$^{2,3}$}

\affiliation{
$^1$Institut f\"{u}r Theoretische Physik, Karl-Franzens-Universit\"{a}t Graz,\\ 
Universit\"{a}tsplatz 5, A-8010 Graz, Austria\\
$^2$Condensed Matter Theory Group, Department of Physics, Uppsala University,\\
BOX 530, S-751 21, Uppsala, Sweden\\
$^3$Applied Materials Physics, Department of Materials Science and Engineering,\\
Royal Institute of Technology, Brinellvagen 23, SE-100 44, Stockholm, Sweden
}

\date{\today}

\begin{abstract}
The lattice dynamical properties of Li$_x$BC are calculated for several values of $x$ using density functional perturbation theory.
We find that the electron-phonon coupling parameter $\lambda$ increases monotonically with decreasing $x$ to a maximum value of 1.4 for
$x=0.125$ owing to the increasing radius of multiply-nested Fermi surface cylinders. The B-C bond-stretching phonon modes have
frequencies which are 28\% higher than the equivalent modes in MgB$_2$. This combination results in a $T_c$ of about 65 K for $x=0.5$.
\end{abstract}

\maketitle
It has been suggested recently that hole doped LiBC may have a superconducting transition temperature similar to or
exceeding that of the isovalent compound MgB$_2$ \cite{Rosner02,Ravindran01}. Experimental confirmation of this would be
of great significance since MgB$_2$ already has an exceptionally high transition temperature of $\sim 40$ K.
Since the discovery of superconductivity in MgB$_2$ \cite{Nagamatsu01} there has been a flurry of interest in it and compounds with
similar features, namely, layered crystals with weak interlayer interactions, a corresponding high density of states at the Fermi
energy and strong in-plane covalent bonds which are partially filled
\cite{An01,Choi02,Rosner02b,Bohnen01,Chen01,Okatov01,Shein01,Kurmaev02,Joas02,Profeta02b,Singh01,Belashchenko01,Yildirim01,Kortus01}. 
LiBC has a hexagonal structure with alternating lithium and boron-carbon planes.
Although it is semi-conducting owing to the alternating B-C atoms within each plane \cite{Rosner02},
the addition of holes, likely to be accomplished by the removal of Li atoms, cause it to become metallic and thus potentially superconducting.
To date no such effect has been observed experimentally which is the motivation for this article in which we report a first-principles
calculation of the electron-phonon coupling in Li$_x$BC.

\begin{figure}
\centerline{\includegraphics[scale=1,angle=0,clip = true]{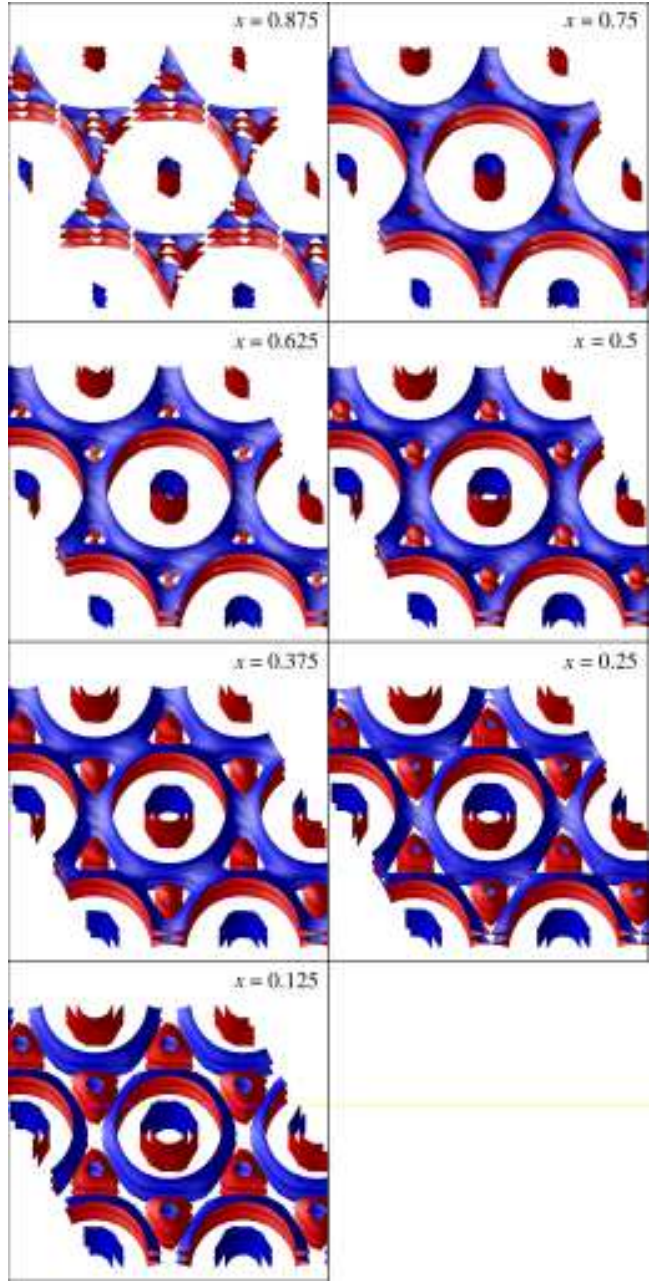}}
\caption{Fermi surfaces viewed along the $c$ axis for various hole dopings.}
\label{fsurf}
\end{figure}

To obtain accurate results we employ a plane wave method with the analytic pseudopotentials of Hartwigsen, Goedecker and Hutter using
the parameters quoted in their paper \cite{Hartwigsen98}. A  kinetic energy cut-off
criterion of 24 Hartree was placed on the plane waves to ensure good convergence in terms of basis size.
The density functional linear response method of Giannozzi {\it et al.} \cite{Giannozzi91} was used to calculate the phonon spectra and corresponding
self-consistent linearized potentials within the local density approximation \cite{Ceperley80,Perdew81}. Dynamical matrices
and linearized potentials were calculated in the irreducible Brillouin zone on a $4 \times 4 \times 2$ $q$-point mesh using the ground-state
wave functions obtained on a $8 \times 8 \times 4$ $k$-point mesh. To incorporate hole doping we removed the appropriate number of electrons
from the unit cell and added a positive uniform background to maintain charge neutrality. This approximation is justified from the results of
Rosner, Kitaigorodsky and Pickett (RKP) \cite{Rosner02} who noted an almost rigid shift in the Fermi level with little
renormalization of the electronic density of states upon
removal of lithium atoms. The phonon spectra and electron-phonon coupling results were calculated for $x=0.125$, 0.25, 0.375, 0.5, 0.625,
0.75 and 0.875. We also performed a ground-state calculation on an even denser $16 \times 16 \times 8$ $k$-point mesh for producing the
electron-phonon coupling matrices as well the Fermi surface plots in Fig. \ref{fsurf}.

\begin{figure}
\centerline{\includegraphics[scale=0.5,angle=0, clip = true]{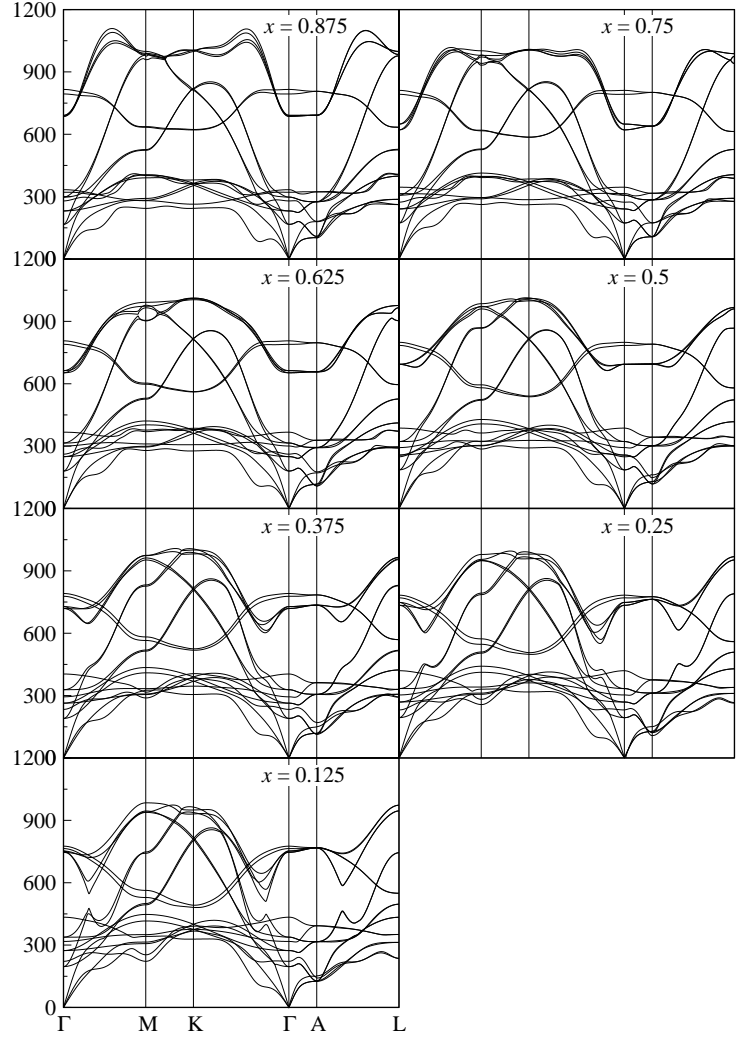}}
\caption{Calculated phonon dispersions along high symmetry directions for various hole dopings. Frequency units are cm$^{-1}$.}
\label{phdisp}
\end{figure}

In Fig. \ref{phdisp} we plot the calculated phonon dispersions along high symmetry lines for each doping level. As is evidenced from
the lack of singularities in the phonon curves, the material is found to be stable for all doping concentrations within the approximation of
a uniform background. The most noteworthy feature in the spectra is the abrupt dip in the topmost branches between the $\Gamma-M$,
$K-\Gamma$ and $A-L$ points. As with MgB$_2$, these branches have the $E_{2g}$ symmetry character and the dip arises from the exceptionally strong
nesting between concentric cylinders \cite{Kortus01} of the Fermi surfaces which are oriented in the direction of the
$c$-axis as shown in Fig. \ref{fsurf}. These nesting modes correspond to in-plane bond stretching between the boron and carbon atoms. With decreasing
$x$ (i.e. increasing doping) the sharpness of the dip becomes less pronounced although the midpoint moves away from the central axis.
Also of interest is the level attraction between certain phonon branches for $x<0.5$. This is reflected in the Fermi surface cylinders which
change from being nearly circular for low doping to hexagonal for high doping. This flattening of the sheets results in additional translational symmetries
in the effective potential. The C-B bond-stretching modes have phonon frequencies around 1000 cm$^{-1}$ which is considerably higher than the B-B
frequencies of MgB2 which are about 725 cm$^{-1}$ \cite{Yildirim01}.

\begin{figure}
\centerline{\includegraphics[scale=0.5,angle=-90, clip = true]{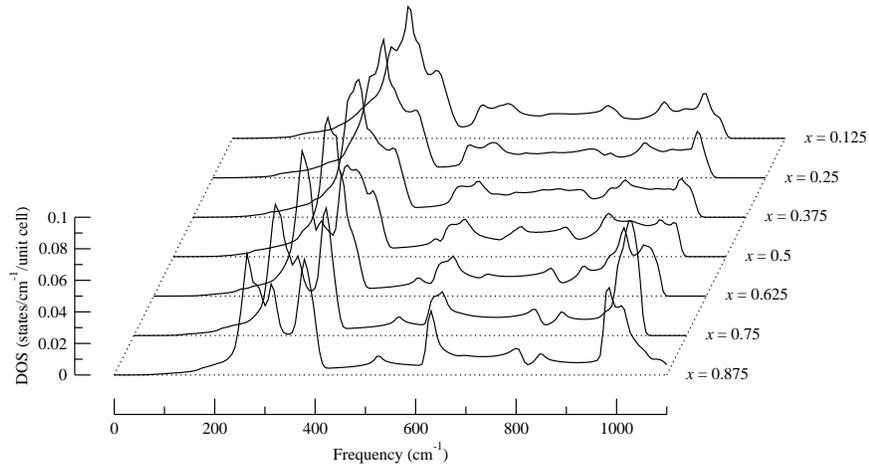}}
\caption{Calculated phonon density of states for various hole dopings.}
\label{phdos}
\end{figure}

The phonon density of states are plotted in Fig. \ref{phdos}. At low doping levels there is a double peak structure between 200 and 400 cm$^{-1}$
as well as peaks at 620 and 1000 cm$^{-1}$. As the doping level is raised the lower peaks merge together and the higher frequency peaks become
much less pronounced. The gap-like structure in the density of states between 400 and 600 cm$^{-1}$ disappears with increased doping.

We now turn to calculation of the superconducting transition temperature. Details of Migdal-Eliashberg theory which
is the basis for these calculations can be found in Ref. \cite{Allen82}. The electron-phonon interaction matrix is determined from the
linearized self-consistent potential $\delta V_{\bf q}^{\rm SCF}$ by
\begin{eqnarray}
g_{{\bf k}+{\bf q}v',{\bf k}v}^{{\bf q}j}= \left\langle
 \phi_{{\bf k}+{\bf q}v'} \left| \sum_i \sqrt{\frac{1}{2M_i \omega_{{\bf q}j}}} {\bf e}_{{\bf q}j;i}
 \cdot \delta V_{{\bf q};i}^{\rm SCF}\right| \phi_{{\bf k}v} \right\rangle,
\end{eqnarray}
where $M_{i}$ is the mass of the $i$'th ion, $\phi_{{\bf k}v}$ is a single-particle state and ${\bf e}_{{\bf q}j;i}$
is the eigenvector of phonon mode ${\bf q}j$ and atom $i$. From this we obtain the phonon linewidth arising from
the electron-phonon interaction by averaging over the Fermi surface
\begin{eqnarray}
\gamma_{{\bf q}j}=2\pi \omega_{{\bf q}j} \sum_{{\bf k}vv'} |g_{{\bf k}+{\bf q}v',{\bf k}v}^{{\bf q}j}|^2
 \delta(\epsilon_{{\bf k}v}-\epsilon_{\rm F})\delta(\epsilon_{{\bf k}+{\bf q}v'}-\epsilon_{\rm F}),
\end{eqnarray}
where $\epsilon_{{\bf k}v}$ is the Kohn-Sham \cite{Kohn65} eigenvalue and $\epsilon_{\rm F}$ is the Fermi energy. The central quantity in
superconductivity theory is the electron-phonon spectral function $\alpha^2 F(\omega)$. This can be written in terms of the phonon
linewidth by
\begin{eqnarray}
\alpha^2 F(\omega)=\frac{1}{2 \pi N_{\rm F}} \sum_{{\bf q}j} \frac{\gamma_{{\bf q}j}}{\omega_{{\bf q}j}}
\delta(\omega - \omega_{{\bf q}j}),
\end{eqnarray}
where $N_{\rm F}$ is the density of states at the Fermi energy.
The total electron-phonon coupling parameter is defined as the first reciprocal moment of $\alpha^2 F(\omega)$
\begin{eqnarray}
\lambda=2 \int_0^{\infty} \frac{\alpha^2 F(\omega)}{\omega} \, d\omega.
\end{eqnarray}
Finally $T_c$ can be estimated from the McMillan equation \cite{McMillan68},
\begin{eqnarray}\label{eqmcmillan}
T_c=\frac{w_{\rm log}}{1.2} {\rm exp} \left( \frac{-1.04 (1+\lambda)}{\lambda-\mu^* (1+0.62\lambda)} \right),
\end{eqnarray}
with $\mu^*$, the Morel-Anderson pseudopotential, representing Coulombic repulsion and
the logarithmic average frequency given by
\begin{eqnarray}
w_{\rm log}=\exp \left\{ \frac{2}{\lambda} \int_{0}^{\infty} \alpha^2 F(\omega) \frac{\ln \omega}{\omega} \right\}.
\end{eqnarray}

\begin{figure}
\centerline{\includegraphics[scale=0.5,angle=-90, clip = true]{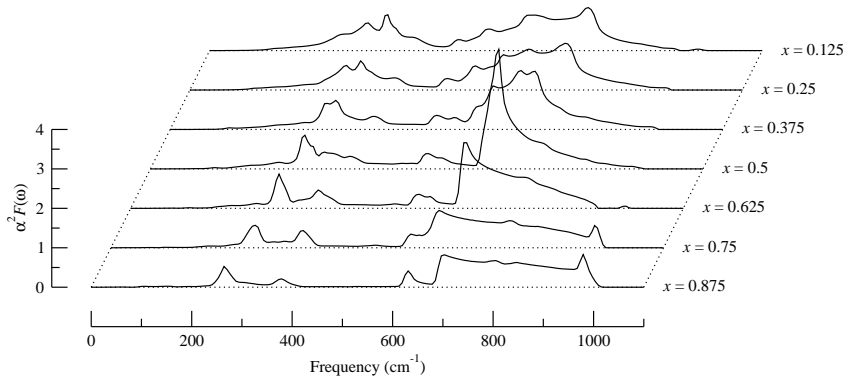}}
\caption{Calculated electron-phonon spectral function for various hole dopings.}
\label{alpha2f}
\end{figure}

\begin{figure}
\centerline{\includegraphics[scale=0.5,angle=0, clip = true]{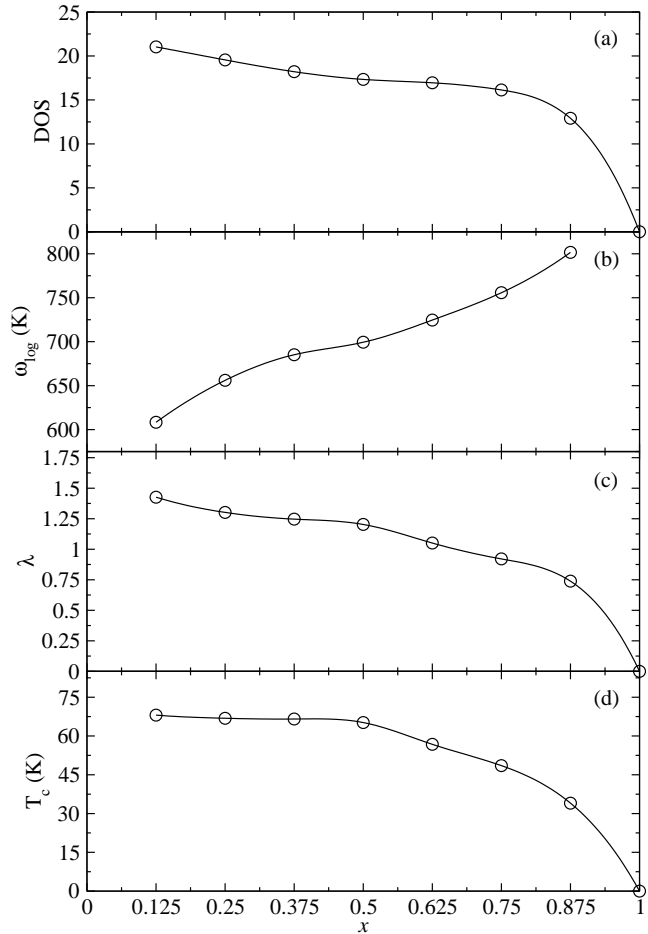}}
\caption{Properties calculated as a function of Li concentration $x$: density of states at Fermi energy in units of
states/spin/Hartree/unit cell (a), logarithmic average frequency (b), electron-phonon coupling parameter $\lambda$ (c),
and superconducting transition temperature from the McMillan formula (d). The circles represent calculated points
and the interpolating lines are cubic splines.}
\label{multi}
\end{figure}

The three main components in determining $T_c$ are therefore the density of states
at the Fermi energy $N_{\rm F}$, the logarithmic average frequency $w_{\rm log}$ and the strength of the electron-phonon coupling
on the Fermi surface. By plotting $\alpha^2 F(\omega)$ we can gauge the relative strength of the coupling. This is done in Fig. \ref{alpha2f}.
For most phonon-mediated superconductors the phonon density of states and $\alpha^2 F(\omega)$ differ only slightly \cite{Savrasov96}.
This is not the case here: the modes above 600 cm$^{-1}$ which dominate in $\alpha^2 F(\omega)$ are suppressed in the phonon density
of states in Fig. \ref{phdos}. This is particularly evident for the peak at $x=0.5$ and $\omega \approx 600$ cm$^{-1}$ which corresponds
to the weakly dispersed $E_{2g}$ branch in the $\Gamma-A$ direction.
We plot in Fig. \ref{multi} $N_{\rm F}$, $w_{\rm log}$, $\lambda$, and $T_c$ using an empirical value for $\mu^*$ of $0.09$ which was
found to reproduce $T_c$ in MgB$_2$ \cite{Rosner02}. The density of states at the Fermi energy increases sharply from $x=1$ to $x=0.75$.
Beyond this $N_{\rm F}$ increases more gradually with a slight upturn at $x=0.5$ and has values between 15 and 21 states/spin/Hartree/unit cell.
As hole concentration is raised $w_{\rm log}$ drops by about 25\% from 802 K at $x=0.875$ to 608 K at $x=0.125$. There is also an
inflection point at $x=0.5$ in this curve.
The total electron-phonon coupling parameter $\lambda$ increases over the whole range to a maximum value of 1.43. This is a direct
result of the increase with $1-x$ of the Fermi surface nesting vector $2k_{\rm F}$ which maps the opposing walls of the cylindrical
sheets shown in Fig. \ref{fsurf}. We again note an inflection point in $\lambda$ for $x=0.5$. From proportionality arguments, RKP
obtained $\lambda \approx 1.5$ for $x=0.5$ which is larger than the 1.20 found here. Nonetheless, this is substantially larger
than $\lambda=0.82$ found in MgB$_2$ \cite{Kong01,Bohnen01,Yildirim01,Liu01}.
Despite the fairly large increase in $\lambda$ over the whole range of $x$, $T_c$ reaches a plateau of 65 K at $x=0.5$.
The reason for this is the drop in $w_{\rm log}$ as a function of $1-x$ which enters as a proportionality in Eq. \ref{eqmcmillan}
as well as the saturation of $N_{\rm F}$ for $x<0.75$. The highest value for $T_c$ is 68 K at $x=0.125$, this is substantially
higher than that observed for MgB$_2$ but lower than the $\sim 100$ K estimated by RKP.

In conclusion, we find from our first-principles calculations that hole-doped LiBC is a superconductor with exceptionally
strong in-plane coupling. As with MgB$_2$ this is due to a small number of phonon modes close to the $\Gamma$ point which scatter electrons
at the Fermi surface from one side of a cylindrical sheet to another. Although the superconducting transition temperature is much higher than
in MgB$_2$, the decrease in the average phonon frequency as the doping concentration is raised limits the potential gain in $T_c$ from
the increase in $\lambda$ and we note that hole doping beyond $x=0.5$ will not result in any substantial improvement $T_c$.
Nonetheless, going by these results, Li$_x$BC may be an impressive superconductor well worth further investigation.

We would like to acknowledge the Swedish Foundation for International Cooperation in Research and
Higher Education (STINT), the Austrian Science Fund (Project P13430) and the EXCITING network (Contract HPRN-CT-2002-00317)
for financial support.


\end{document}